\documentclass[sigconf]{acmart}

\AtBeginDocument{%
  }

\usepackage{amsmath}   
\usepackage{amssymb}   

\usepackage{graphicx}  
\usepackage{tikz}      
\usepackage{float}     

\usepackage{xcolor}    
\usepackage{colortbl}  

\usepackage{tcolorbox} 
\usepackage{mdframed}  

\usepackage{zi4} 
\usepackage{lipsum}      

\usepackage{multirow}   
\usepackage{booktabs}   
\usepackage{array}      
\usepackage{adjustbox}  
\usepackage{balance}

\usepackage{fancyvrb}   
\copyrightyear{2025}
\acmYear{2025}
\setcopyright{acmlicensed}
\acmConference[SIGIR-AP '25] {Proceedings of the 2025 Annual International ACM SIGIR Conference on Research and Development in Information Retrieval in the Asia Pacific Region}{December 7--31, 2025}{Dublin, Ireland.}
\acmBooktitle{Proceedings of the 2025 Annual International ACM SIGIR Conference on Research and Development in Information Retrieval in the Asia Pacific Region (SIGIR-AP '25), December 7--10, 2015, Xi'an, China}
\acmISBN{979-8-4007-2218-9/2025/12}
\acmDOI{10.1145/XXXXXX.XXXXXX}

\author{Haowei Liu}
\affiliation{
  \institution{Santa Clara University}
  \city{Santa Clara}
  \state{CA}
  \country{USA}
}
\email{hliu6@scu.edu}

\author{Xuyang Wu}
\affiliation{
  \institution{Santa Clara University}
  \city{Santa Clara}
  \state{CA}
  \country{USA}
}
\email{xwu5@scu.edu}

\author{Guohao Sun}
\affiliation{
  \institution{Rochester Institute of Technology}
  \city{Rochester}
  \state{NY}
  \country{USA}
}
\email{gs4288@rit.edu}

\author{Hsin-Tai Wu}
\affiliation{
  \institution{DOCOMO Innovations}
  \city{Sunnyvale}
  \state{CA}
  \country{USA}
}
\email{hwu@docomoinnovations.com}

\author{Zhiqiang Tao}
\affiliation{
  \institution{Rochester Institute of Technology}
  \city{Rochester}
  \state{NY}
  \country{USA}
}
\email{zhiqiang.tao@rit.edu}

\author{Yi Fang}
\authornote{Corresponding author.}
\affiliation{
  \institution{Santa Clara University}
  \city{Santa Clara}
  \state{CA}
  \country{USA}
}
\email{yfang@scu.edu}

\begin{document}

\title{RaCT: Ranking-aware Chain-of-Thought Optimization for LLMs}

\renewcommand{\shortauthors}{Haowei Liu et al.}

\begin{abstract}
In information retrieval, large language models (LLMs) have demonstrated remarkable potential in text reranking tasks by leveraging their sophisticated natural language understanding and advanced reasoning capabilities. However, conventional supervised fine-tuning approaches for specializing LLMs in ranking tasks often lead to significant degradation of the models' general-purpose abilities. To address this fundamental challenge, this paper presents a novel methodology that strategically combines Chain-of-Thought (CoT) prompting techniques with an innovative two-stage training pipeline consisting of Supervised Fine-Tuning followed by Ranking Preference Optimization (SFT-RPO). The Chain-of-Thought prompting component encourages models to explicitly articulate their reasoning process during ranking decisions, creating a transparent pathway from query-document analysis to final ranking scores while maintaining analytical capabilities throughout fine-tuning. Extensive experimental evaluations on the TREC Deep Learning datasets demonstrate that our proposed method achieves superior performance compared to existing state-of-the-art models, including RankZephyr, showing consistent improvements across multiple evaluation metrics such as normalized Discounted Cumulative Gain (nDCG). Most significantly, comprehensive assessments on the Massive Multitask Language Understanding (MMLU) benchmark reveal that our method successfully maintains robust performance across diverse reasoning tasks, providing strong empirical evidence for effective retention of general-purpose capabilities through strategic fine-tuning while achieving specialized performance improvements in text reranking.
\end{abstract}




\maketitle

\section{Introduction}

Text reranking is a vital task in information retrieval \cite{DBLP:journals/ftir/Liu09, DBLP:conf/wsdm/Hasanain18}, crucial for search engines \cite{DBLP:journals/corr/abs-2404-16924}, conversational AI \cite{DBLP:conf/bea/BeckerPVW12}, and recommendation systems \cite{DBLP:journals/corr/abs-2407-01712}. Large language models (LLMs) excel in reranking due to their reasoning and human-like thinking capabilities \cite{brown2020languagemodelsfewshotlearners}, enabling them to handle complex queries and ambiguous contexts. This paradigm shift represents a departure from traditional term-based matching approaches toward more nuanced semantic understanding, enabling systems to capture complex query-document relationships that were previously difficult to model effectively.

\begin{figure}[htbp]
    \centering
    \includegraphics[width=\columnwidth]{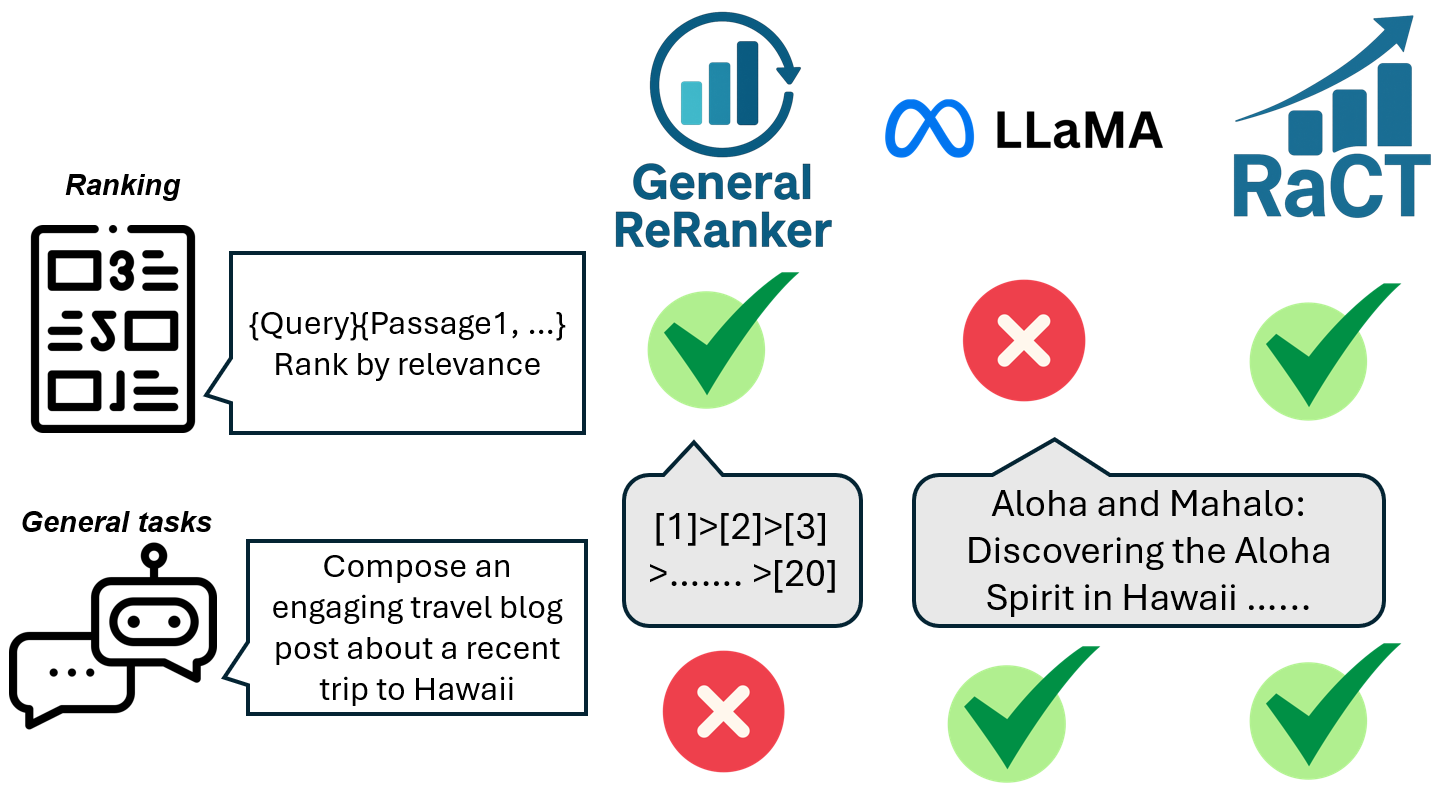} 
    \caption{General text reranking LLMs excel at ranking tasks but struggle with general tasks such as open-ended text generation. General LLMs (e.g., LLaMA) perform well on diverse tasks but lack strong ranking capabilities. RaCT effectively integrates both, achieving high performance in ranking while maintaining general language generation abilities.}
    \label{fig:llm example}
\end{figure}

The emergence of LLM-based reranking, pioneered by models like RankGPT \cite{achiam2023gpt, sun2023chatgpt}, has established that properly instructed language models can achieve competitive or superior performance to traditional neural rankers without task-specific training. This advancement represents a fundamental shift from statistical ranking methods to reasoning-based approaches, enabling systems to move beyond keyword matching toward nuanced semantic understanding of query-document relationships. The impact extends across diverse applications, from web search and enterprise retrieval systems that can now understand implicit user intent, to conversational AI platforms that can contextually rank information within dialogue, and recommendation systems that leverage natural language understanding for more accurate content suggestion.

However, this remarkable progress has revealed a critical and underexplored challenge: \emph{the fundamental trade-off between ranking specialization and general language modeling capabilities}. While existing approaches like RankVicuna \cite{pradeep2023rankvicuna} and RankZephyr \cite{pradeep2023rankzephyr} achieve state-of-the-art ranking performance through supervised fine-tuning (SFT), they suffer from severe forgetting that dramatically degrades their ability to perform general language understanding and generation tasks. This degradation is not merely a minor side effect—it represents a fundamental limitation that restricts the practical deployment of these systems in real-world applications.

As illustrated in \ref{fig:llm example}, models optimized exclusively for ranking often produce incomplete, incoherent, or entirely irrelevant outputs when asked to perform everyday language generation tasks, such as writing a travel blog, composing professional correspondence, or engaging in natural dialogue. Existing specialized ranking models exhibit significant degradation in general language capabilities, with some losing the ability to perform basic text generation and comprehension tasks that were well within the capabilities of their base models before fine-tuning.

In this study, we introduce a novel LLM reranking algorithm, namely Ranking-aware Chain-of-Thought Optimization(RaCT), a novel approach that combines the reasoning capabilities of Chain-of-Thought (CoT) prompting \cite{wei2023chainofthoughtpromptingelicitsreasoning} with a carefully designed two-stage training pipeline. This design is consistent with survey evidence that step-wise, feedback-guided reasoning helps LLMs handle complex problems; we adapt that idea to listwise ranking \cite{wei2025surveyfeedbackbasedmultistepreasoning}. Our key insight is that by formulating ranking as an explicit reasoning process through CoT prompting (see Fig.~\ref{fig:user prompt}), we can enhance ranking performance while preserving and potentially even leveraging the model's inherent language modeling abilities.

The intuition behind our approach is straightforward: instead of asking the model to output the complete ranked list all at once, RaCT uses Chain-of-Thought prompting to guide the model through an incremental ranking process. The model first selects the most relevant document, then builds a ranking of the top two documents, then the top three, and so on, step by step until the complete ranking is formed. This incremental approach not only improves ranking accuracy but also maintains the model's natural language reasoning abilities throughout the fine-tuning process.

RaCT implements a carefully orchestrated two-stage optimization strategy: (1) CoT-guided Supervised Fine-Tuning, where the model learns to perform step-by-step passage ranking through structured reasoning chains, and (2) Ranking Preference Optimization (RPO), a novel adaptation of Direct Preference Optimization \cite{rafailov2024direct} that uses overlapping ranking orders as reward signals to further refine the model's ranking decisions while maintaining consistency in its reasoning process.

We summarize the contributions of this work as follows:
\begin{itemize}
    \item To the best of our knowledge, this is the first research study to investigate the trade-off between ranking utility and language modeling for the recent emerging LLM re-rankers. 
    \item We propose a novel chain-of-thoughts instruction (reranking) tuning algorithm -- RaCT -- that enables LLMs to rank passages based on relevance step by step, and RPO to enable further ranking preference optimization. 
    \item Empirical evidence on three ranking benchmarks, TREC Deep Learning Tracks \cite{craswell2021overviewtrec2020deep, craswell2020overview}, BEIR~\cite{DBLP:journals/corr/abs-2104-08663}, and BRIGHT~\cite{su2025brightrealisticchallengingbenchmark}, and the Massive Multitask Language Understanding (MMLU) benchmark~\cite{hendrycks2020measuring} suggests that our approach achieves state-of-the-art LLM re-ranking performance with the preservation of intrinsic language modeling capabilities.
\end{itemize}

\section{Related Work}
Our work builds upon several interconnected research areas in large language models and information retrieval. We organize the related work into four key areas that directly inform our approach.

\subsection{LLM-based Text Reranking}
The use of Large Language Models for reranking has advanced quickly, giving rise to three paradigms. Pointwise methods evaluate each document independently but often encounter calibration issues with generation-only APIs. Pairwise methods, such as PRP \cite{qin2024largelanguagemodelseffective}, reduce task complexity by comparing pairs, with FLAN-UL2 (20B) achieving performance comparable to GPT-4 on TREC-DL. Listwise approaches, however, have become especially effective, as they exploit LLMs’ ability to reason across multiple documents and output coherent ranked sequences.

RankGPT \cite{sun2023chatgpt} demonstrated that well-instructed LLMs can directly generate reordered lists through zero-shot permutation with sliding window strategies, showing LLMs are naturally suited to listwise ranking. LRL \cite{ma2023zero} confirmed this by producing document identifier sequences that surpassed pointwise baselines on several TREC benchmarks.

Building on this, subsequent work targeted reproducibility and efficiency. RankVicuna \cite{pradeep2023rankvicuna} distilled RankGPT-3.5 into a 7B model, while RankZephyr \cite{pradeep2023rankzephyr} improved performance by training on outputs from both RankGPT-3.5 and GPT-4. Other advances include instruction distillation \cite{sun2023chatgpt} for compressing pairwise into efficient pointwise ranking, setwise decomposition \cite{qin2024largelanguagemodelseffective} for partitioning tasks, GPT-independent listwise models \cite{zhang2023rankwithoutgptbuildinggptindependentlistwise}, and LiT5 \cite{tamber2023scalingdownlittingup}, which uses Fusion-in-Decoder for efficient listwise reranking with smaller models. While these methods enhance effectiveness and efficiency, they mainly optimize ranking quality and often overlook preserving general language modeling abilities.

Beyond performance, fairness has also been examined. Meta-learning approaches with curriculum design have been proposed to encourage fairness in outputs \cite{10.1109/TKDE.2024.3377644}, though these remain orthogonal to our focus on maintaining broader language modeling capabilities.

\subsection{Chain-of-Thoughts in Information Retrieval}

While Chain-of-Thought prompting has been extensively studied for mathematical and logical reasoning tasks \cite{wei2023chainofthoughtpromptingelicitsreasoning}, its application to information retrieval represents an emerging and promising direction. IRCoT \cite{trivedi2023interleavingretrievalchainofthoughtreasoning} pioneered the integration of retrieval with CoT reasoning for knowledge-intensive multi-step questions, creating a feedback loop that improved both retrieval and downstream QA performance.

Subsequent work found that explicit reasoning can be elicited even without few-shot examples: simply appending a phrase like “Let’s think step by step” to the query (Zero-Shot CoT) induces the model to produce a chain of thought and dramatically boosts zero-shot reasoning accuracy \cite{kojima2023largelanguagemodelszeroshot}. Such findings suggest that many LLMs have latent reasoning ability that can be unlocked with the right prompts. CoT reasoning has since been applied to a wide array of tasks – from commonsense QA to multi-step knowledge tests \cite{yao2023treethoughtsdeliberateproblem, li2023machinelearningmietronics, wang2023selfconsistencyimproveschainthought, zhou2023leasttomostpromptingenablescomplex}. Even on broad academic benchmarks like MMLU, which cover 57 diverse subjects, using CoT-style reasoning leads to better performance on the more challenging questions that require complex deduction. Additionally, CoT-RAG \cite{li2025cotragintegratingchainthought} integrated knowledge graphs with a learnable case-aware retrieval-augmented generation (RAG) framework, achieving accuracy gains ranging from 4.0\% to 44.3\% across nine reasoning datasets. This work highlights how structured knowledge integration can enhance the reliability of reasoning in retrieval contexts. Beyond textual reasoning, multimodal systems have also shown the value of step-wise strategies. SQ-LLaVA \cite{sun2024sqllavaselfquestioninglargevisionlanguage} introduced a self-questioning framework for vision-language assistants, where models recursively generate and answer intermediate questions before finalizing outputs. This mirrors Chain-of-Thought prompting by explicitly structuring reasoning steps, reinforcing that step-wise reasoning can improve both accuracy and interpretability across tasks.

However, existing work has primarily focused on query processing and retrieval rather than the ranking phase itself. Our work represents the first systematic application of Chain-of-Thought prompting specifically to document reranking, where the reasoning process directly guides ranking decisions.

\subsection{Preference Optimization for LLMs}

The development of preference optimization methods for large language models has evolved from complex reinforcement learning frameworks toward more direct and efficient approaches. Reinforcement Learning from Human Feedback (RLHF) established the foundation for aligning language models with human preferences through reward model training and policy optimization \cite{ouyang2022traininglanguagemodelsfollow}. However, RLHF requires complex multi-stage training procedures involving reward model fitting, policy optimization through reinforcement learning, and careful hyperparameter tuning to maintain training stability.

Direct Preference Optimization (DPO) \cite{rafailov2024direct} introduced a paradigm shift by eliminating the explicit reward modeling stage and directly optimizing the policy using preference data. DPO leverages the mathematical relationship between optimal policies and reward functions under the Bradley-Terry preference model, enabling preference optimization through a simple classification objective. This approach demonstrates superior stability and computational efficiency compared to traditional RLHF methods while achieving comparable or improved performance across various tasks.

Preference Ranking Optimization (PRO) \cite{song2024preferencerankingoptimizationhuman} addresses limitations of RLHF by extending pairwise preference comparisons to accommodate human preference rankings of any length. Rather than relying on the complex reward modeling and PPO \cite{schulman2017proximalpolicyoptimizationalgorithms} training pipeline of traditional RLHF, PRO directly fine-tunes LLMs using supervised learning on human preference rankings. The method takes multiple responses generated by an LLM and aligns the model's probability ranking of these responses with human preference rankings through iterative contrast learning. PRO transforms the human alignment task into matching probability distributions rather than requiring explicit reward model training, achieving comparable results to ChatGPT while maintaining the stability of supervised fine-tuning.

Our work contributes to this research direction by introducing Ranking Preference Optimization (RPO), which addresses the specific requirements of Chain-of-Thought ranking processes. RPO utilizes overlapping ranking sequences as preference signals, enabling the optimization of both ranking decisions and reasoning consistency within a unified framework.

\subsection{Positioning of Our Work}
RaCT emerges at the intersection of LLM reranking, Chain-of-Thought reasoning, and preference optimization. Prior ranking models emphasize effectiveness but often compromise general language abilities, while CoT applications to IR remain limited to query processing. We address this gap by combining CoT with capability-preserving training, ensuring that reasoning can directly guide the reranking process without eroding broader LLM utility.

Our contribution is to frame ranking as an explicit reasoning task and refine it through Ranking Preference Optimization (RPO). This integration balances specialization with generalization, enabling RaCT to deliver strong ranking performance while retaining explainability and the versatility required for diverse language tasks.

\begin{figure}[t]
    \centering
    \begin{mdframed}[
        linecolor=black!60,  
        linewidth=1pt,        
        roundcorner=10pt,     
        backgroundcolor=gray!5,  
        shadow=true,          
        shadowsize=5pt,       
        shadowcolor=black!40, 
        skipabove=10pt,       
        skipbelow=10pt,       
        innertopmargin=10pt,  
        innerbottommargin=10pt, 
        innerleftmargin=10pt, 
        innerrightmargin=10pt 
    ]
    \small
    \textbf{USER:} I will provide you with \{num\} passages, each indicated by a numerical identifier []. Rank the passages based on their relevance to the search query: \{query\}. \\
    \texttt{[1]} \{passage 1\} \\
    \texttt{[2]} \{passage 2\} \\
    \texttt{...} \\
    \texttt{[\{num\}]} \{passage \{num\}\} \\
    \textbf{Search Query:} \{query\}. \\
    Rank the \{num\} passages by selecting the most relevant passage at each step from the remaining passages. After choosing the most relevant passage, remove it from the pool and continue ranking until all passages are ordered.\\
    \textbf{Instructions:} \\
    Start with the most relevant passage and select it from the full list. \\
    For each following step, pick the most relevant passage from the remaining passages only.\\
    List the selected passages by their identifiers at each step, one after the other, until all passages are ranked. \\
    \textbf{Example Output:}\\
    Step 1: [4] \\
    Step 2: [4, 2] \\
    Step 3: [4, 2, 3] \\
    \texttt{...} \\
    step \{num\}: [4,2,3,15,...,14]\\
    Final Answer: [4, 2, 3,..., 14]\\
    Only respond with each step and the final answer, ensuring each passage is included once and ranked in descending relevance.
    \end{mdframed}

    \caption{RaCT Chain-of-Thought (CoT) reranking prompt guiding the model to rank passages based on relevance to a query iteratively. The prompt ensures step-by-step selection, removal, and ordering of passages, with an example illustrating the expected output format.}\vspace{-0.2cm}
    \label{fig:user prompt}
\end{figure}

\section{Methodology}
\begin{figure*}[htbp]
    \centering
    \includegraphics[width=\textwidth]{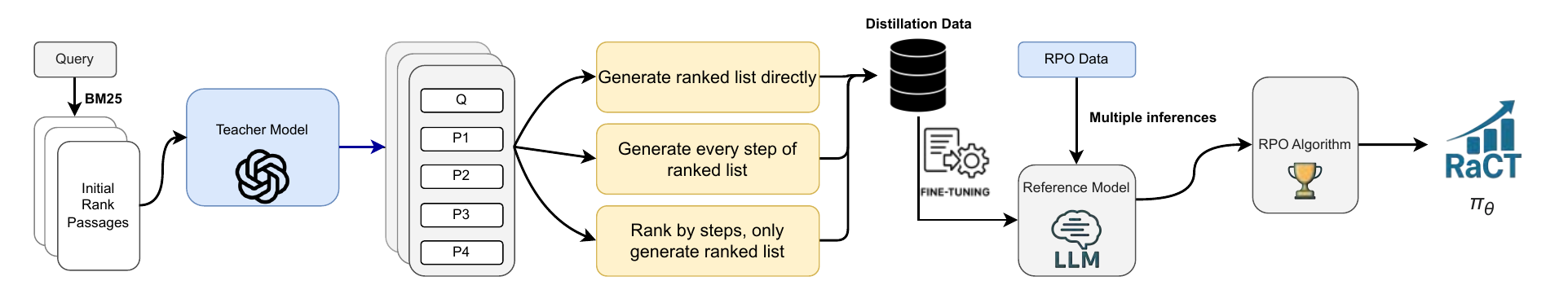}
    \caption{RaCT two-stage training pipeline. BM25 retrieves initial candidate passages given a query. A teacher model (e.g., GPT-4) generates ranking supervision in three formats. These form the distillation dataset used to fine-tune a LLaMA-based model. In the second stage, the fine-tuned model generates multiple ranking trajectories on held-out prompts, which are evaluated against reference model predictions using the Ranking Preference Optimization (RPO) algorithm. The resulting preference-labeled data guides final optimization of RaCT.}
    \label{fig:framework}
\end{figure*}

\subsection{CoT Reranking Prompt}

RaCT formulates listwise reranking as a chain-of-thought (CoT) reasoning task, where the model iteratively selects the most relevant passage until all are ranked. The prompt (Figure~\ref{fig:user prompt}) guides LLaMA to output ranked indices given a query and a set of passages. This formulation encourages the model to perform stepwise relevance judgments, conditioning each decision on the previously selected passages.

This design builds on prior zero-shot reranking methods such as RankVicuna and RankZephyr~\cite{pradeep2023rankvicuna, pradeep2023rankzephyr, liu2024leveraging}, which use template-based prompts to optimize ranking metrics like nDCG~\cite{DBLP:journals/tois/JarvelinK02}. However, RaCT differs in that it explicitly structures the ranking process as a sequential reasoning task, making it amenable to both supervised fine-tuning and preference-based optimization. By modeling the ranking decision as a chain of incremental steps, the prompt design supports interpretability, enables intermediate supervision, and improves robustness across different inference scenarios.

\subsection{Data Construction}
We train RaCT on 40k examples from \citet{pradeep2023rankzephyr}, where each instance consists of a query and 20 BM25-retrieved passages labeled by RankGPT$_{3.5}$ or RankGPT$_{4}$. To support different styles of reasoning and output behavior, the dataset includes three prompt formats: (1) standard RankGPT-style prompts with direct final list output, (2) RaCT CoT prompts with explicit step-by-step reasoning, and (3) RaCT CoT prompts where the model outputs only the final list while reasoning implicitly.

We split 90\% of the dataset for supervised CoT fine-tuning and 10\% for Ranking Preference Optimization (RPO), referred to respectively as CoT tuning data and RPO data. During supervised training, the model is optimized via Maximum Likelihood Estimation to predict the next passage in the ranked sequence, conditioned on the query and the current prefix. This approach provides finer-grained learning signals than list-level supervision, reinforcing both relevance estimation and consistency in selection.

The dataset is derived from MS MARCO and aligned with standard information retrieval benchmarks, enabling evaluation on diverse reranking tasks. Supervision is obtained from high-quality teacher models rather than human annotators, ensuring scalable and consistent label generation. This allows the model to learn from strong relevance signals while avoiding inconsistencies often seen in manual annotation. The dataset is publicly available.\footnote{\url{https://huggingface.co/datasets/castorini/rank_zephyr_training_data}} This facilitates reproducibility and supports further research on instruction tuning and reasoning-guided reranking.

\subsection{RaCT: Two-Stage Training}

RaCT employs a two-stage training pipeline to specialize large language models for reranking while preserving their general reasoning capabilities. The first stage focuses on teaching the model to perform stepwise ranking via supervised fine-tuning, while the second stage refines these capabilities using Ranking Preference Optimization (RPO), a preference-based learning approach designed to improve consistency and robustness.

\paragraph{Stage 1: Chain-of-Thought Supervised Fine-Tuning.} 

As shown in Figure~\ref{fig:framework}, the first stage uses the CoT tuning subset of the dataset, which contains a mixture of prompt formats to promote generalization across instructional styles. Importantly, all training examples are generated using teacher models (RankGPT$_{3.5}$ and RankGPT$_{4}$), and no human-labeled supervision is involved. This enables a fully zero-shot pipeline in terms of human annotation, while still benefiting from high-quality supervision.

Before fine-tuning, we observe that the pretrained LLaMA3.1-8B-Instruct model~\cite{grattafiori2024llama3herdmodels} fails to perform meaningful CoT-based reranking. It tends to replicate the format of prompt examples without demonstrating actual relevance-based selection, underlining the necessity of domain-specific instruction tuning. Model weights for the base model are publicly available.\footnote{\url{https://huggingface.co/meta-LLaMA/Meta-LLaMA-3-8B-Instruct}}

We perform full fine-tuning on the 8B-parameter model for three epochs using a batch size of 128, a learning rate of $5 \times 10^{-6}$, and bfloat16 numerical precision. Training is conducted on four NVIDIA A100 80GB GPUs, taking approximately 20 hours in total. We adopt Maximum Likelihood Estimation (MLE) as the training objective, where the model is trained to maximize the likelihood of the correct passage being selected at each step, conditioned on the query and the prior selection history. This not only improves the model’s step-level ranking ability but also preserves its reasoning structure across intermediate steps.

This stage ensures that the model acquires the ability to decompose ranking into a sequence of coherent and contextually relevant decisions, providing the foundation for downstream preference optimization.

\begin{figure}[ht]
\centering
\includegraphics[width=1\linewidth]{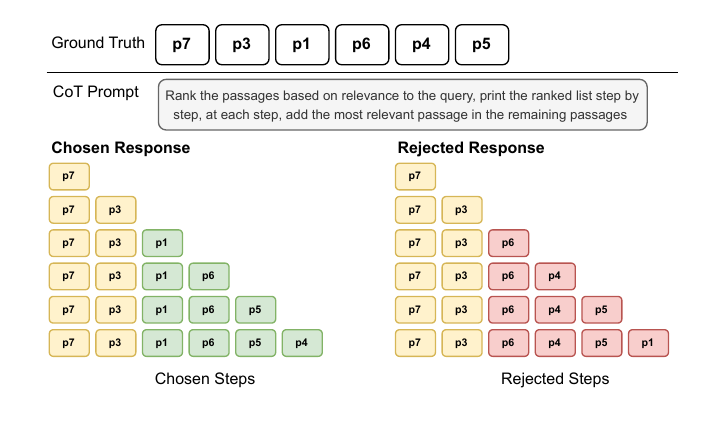}
\caption{RPO decomposes each predicted ranking into a shared prefix ($s_o$), preferred steps ($s_w$), and rejected steps ($s_l$) based on overlap with the ground truth. The model is trained to favor $s_w$ over $s_l$, enhancing stepwise coherence and alignment without using a reward model.}
\label{fig:rpo}
\end{figure}
\paragraph{Stage 2: Ranking Preference Optimization.} 

Following CoT tuning, we apply Ranking Preference Optimization (RPO) to enhance the model’s coherence and preference alignment in stepwise reasoning. In this stage, the trained model generates three candidate rankings $y$ for each prompt $x$ from the RPO subset of the dataset. These predictions are compared against the teacher-provided target to determine overlap-based agreement.

Each final ranking $y = \{s_1, s_2, \dots, s_n\}$ is decomposed into reasoning steps, where each step $s_k$ is conditioned on the prompt and prior decisions via $\pi(s_k \mid x; s_{1:k-1})$. To construct the preference data, we compute the contiguous overlap prefix ($s_o$) shared with the ground-truth sequence. Once divergence occurs, subsequent steps are categorized into chosen ($s_w$) and rejected ($s_l$) completions. This produces a preference-labeled dataset in the form of $(x, s_w, s_l, s_o)$.

To optimize the model, we use a preference-based loss that maximizes the log-probability of preferred steps over rejected ones, scaled by a temperature parameter $\beta$ and regularized using reference model probabilities $\pi_{\text{ref}}$:

\begin{small}
\begin{equation*}
\begin{split}
    \mathcal{L}(\theta) = 
    -\mathbb{E}_{(x,s_{w},s_{l},s_o)\sim D} \Big[
    \log \sigma \Big(
    & \beta \log \frac{\pi_\theta(s_{w} \mid x; s_o)}{\pi_{ref}(s_{w} \mid x; s_o)} \\
    & - \beta \log \frac{\pi_\theta(s_{l} \mid x; s_o)}{\pi_{ref}(s_{l} \mid x; s_o)} 
    \Big) \Big]\,.
\end{split}
\end{equation*}
\end{small}

This formulation allows stable optimization while reinforcing the model’s ability to prefer correct ranking continuations in the context of prior decisions. It avoids the complexity of reinforcement learning frameworks and removes the need for training a separate reward model.

RPO training is conducted for one epoch on four NVIDIA A100 80GB GPUs, requiring approximately 3 hours. Empirically, this stage improves both ranking accuracy and reasoning consistency, particularly under distribution shifts and prompt variations.

Together, these two stages form a coherent optimization strategy: Stage 1 imparts step-level reasoning structure through supervised signals, and Stage 2 aligns these decisions with preference-based objectives to enhance robustness and performance.

\section{Experiments}
\subsection{Research Questions}
For future analysis, we formulate several research questions to ensure that RaCT could effectively address the challenges:

-\textbf{RQ1: }Does the CoT tuning improve the step-by-step relevance ranking of passages compared to traditional ranking methods?

-\textbf{RQ2: }How does our model's performance on text reranking tasks compare with existing models like RankZephyr and RankGPT${_4}$ across various datasets and task settings?

-\textbf{RQ3: }Does introducing RPO improve the performance and robustness of our model?

\subsection{Baseline Selection}
To illustrate the efficacy of RaCT, we select a comprehensive set of baselines spanning traditional retrieval methods, zero-shot LLMs, and fine-tuned LLM-based rerankers. We begin with BM25 \cite{10.1561/1500000019}, a widely used unsupervised sparse retriever that serves as a classical baseline. For zero-shot LLMs, we include a set of publicly available instruction-tuned models—Gemma-7B \cite{team2024gemma}, LLaMA3.1-8B-Instruct \cite{grattafiori2024llama3herdmodels}, and DeepSeek-R1-Distill-Llama-8B \cite{deepseekai2025deepseekr1incentivizingreasoningcapability}—as well as RankGPT$_{4}$\footnote{gpt-4-0613}, which serves dual roles: it is both a strong proprietary LLM known for robust general reasoning capabilities and the teacher model.

We further include RankVicuna and RankZephyr, two fine-tuned open-source rerankers trained with data distilled from RankGPT$_3.5$ and RankGPT$_4$ under a supervised listwise setup. To isolate the effect of our proposed CoT reranking prompt, we introduce a new variant that follows the same training setup, data, and objective as RankZephyr and RankVicuna, but uses LLaMA3.1-8B-Instruct as the base model. We denote this variant as \emph{LLaMA3.1 + SFT}. This controlled comparison allows us to assess the contribution of our CoT-based prompt design independent of model scale, architecture, or data variation.

\begin{table*}[ht]
\centering
\begin{tabular}{lccccccc}
\toprule
\textbf{} & \multicolumn{2}{c}{\textbf{Source}} & \multicolumn{2}{c}{\textbf{MSv1}} & \multicolumn{2}{c}{\textbf{MSv2}} & \textbf{MMLU}\\
\textbf{Model} & \textbf{Prev.} & \textbf{Top-$k$} & \textbf{DL19} & \textbf{DL20} & \textbf{DL21} & \textbf{DL22} & \textbf{AVG.} \\
\midrule
\rowcolor{gray!20}
BM25 & None & $|C|$ & 50.6 & 48.0 & 44.6 & 26.9 & - \\
\midrule
\rowcolor{gray!20}
RankGPT$_{4}$ & BM25 & 100 & 75.0 & 70.4 & 68.4 & 50.9 & 86.4 \\
\midrule
Gemma-7B & BM25 & 100 & 53.3 & 53.0 & 44.6 & 26.9 & 64.9\\
LLaMA3.1 & BM25 & 100  & 66.9 & 64.1 & 63.9 & 42.9  & 72.0\\
DeepSeek-R1-8B  & BM25 & 100 &  56.5 & 50.3 & 55.8 & 38.0 & 53.0\\
\midrule
RankZephyr & BM25 & 100 & 74.2 & 70.9 & 60.0 & 40.8 & 0.0\\
RankVicuna & BM25 & 100  & 66.8 & 65.5 & 60.8 & 42.1 & 37.3\\
LLaMA3.1 + SFT & BM25 & 100  & 70.7 & 63.5 & 66.0 & 44.1 & 62.8 \\
\midrule
RaCT (Ours) & BM25 & 100  & \textbf{75.8}$^{\dagger}$ & \textbf{72.0}$^{\dagger}$ & \textbf{70.6}$^{\dagger}$ & \textbf{53.2}$^{\dagger}$ & \textbf{72.0}\\
\bottomrule
\end{tabular}
\caption{
Performance of different models on TREC (nDCG@10 in \%) and MMLU (exact match score). All reranking tasks are based on BM25 retrieval results. \textbf{Bold} indicates the best score. $^{\dagger}$ indicates statistically significant improvement over RankZephyr ($p < 0.05$), based on paired t-test over query-level nDCG@10.
}
\label{tab:mainresults}
\end{table*}

\begin{table*}[ht]
\centering
\begin{tabular}{llcccccccccccccc}
\toprule
\textbf{Model} & \textbf{Training} & \textbf{Avg.} & \textbf{Climate} & \textbf{DBP-} & \textbf{FEVER} & \textbf{FiQA} & \textbf{Hotpot} & \textbf{Robust} & \textbf{NFC-} & \textbf{NQ} & \textbf{Sci-} & \textbf{Sci-} & \textbf{Trec-} \\
\textbf{} & \textbf{Data} & \textbf{} & \textbf{Fever} & \textbf{edia} & \textbf{} & \textbf{} & \textbf{QA} & \textbf{04} & \textbf{orpus} & \textbf{} & \textbf{docs} &  \textbf{fact} & \textbf{COVID}\\
\midrule
BM25 & MS Marco & 40.6 & 16.5 & 31.8 & 65.1 & 23.6 & 63.3 & 40.7 & 32.2 & 30.6 & 14.9 & 67.9 & 59.5 \\
RankVicuna & GPT-3.5 & 43.3 & 16.3 & 43.2 & 62.8 & 23.6 & 73.0 & 40.8 & 33.9 & 41.2 & 14.7 & 68.1 & 58.8 \\
RankZephyr & GPT-3.5+4 & 41.8 & 16.0 & 39.5 & 61.8 & 23.0 & 66.9 & 40.6 & 33.1 & 37.5 & 14.8 & 67.9 & 59.2 \\
RaCT (ours) & GPT-3.5+4 & \textbf{54.9}$^{\dagger}$ & \textbf{23.7}$^{\dagger}$ & \textbf{45.9}$^{\dagger}$ & \textbf{82.3}$^{\dagger}$ & \textbf{43.5}$^{\dagger}$ & \textbf{76.2}$^{\dagger}$ & \textbf{49.2} & \textbf{39.2} & \textbf{60.0} & \textbf{20.7} & \textbf{79.0} & \textbf{84.7} \\
\bottomrule
\end{tabular}
\caption{Performance of rerankers on BEIR (nDCG@10 in \%). Top-100 BM25 results are reranked. \textbf{Bold} = best performance. $^{\dagger}$ indicates statistically significant improvement over RankZephyr ($p < 0.05$).}
\label{tab:beir-rerankers}
\end{table*}
\subsection{Evaluation Benchmarks}

We evaluate ranking capabilities on a combination of standard and diverse benchmarks to comprehensively assess model performance. For primary evaluation, we use the TREC Deep Learning Tracks DL19~\cite{craswell2020overview} and DL20~\cite{craswell2021overviewtrec2020deep}, which are derived from MS MARCO V1~\cite{bajaj2016ms}. DL19 contains 43 queries and DL20 contains 54 queries, with each query associated with 100 candidate passages retrieved using BM25. These benchmarks include human-annotated relevance labels and are widely adopted for assessing document reranking effectiveness in a controlled setting.

To evaluate generalization under domain shift, we include TREC DL21~\cite{craswell2021overview} and DL22~\cite{craswell2022overview}, which are constructed from the MS MARCO V2 corpus. DL21 includes 66 queries and DL22 contains 75 queries, both associated with 100 BM25-retrieved passages per query. These datasets expand the retrieval scope by introducing new queries and a larger, cleaner passage collection, making them more challenging than DL19 and DL20. Their inclusion allows us to assess how well reranking models generalize to unseen distributions while maintaining evaluation consistency.

For zero-shot evaluation across a broad range of domains, we adopt the BEIR benchmark suite~\cite{DBLP:journals/corr/abs-2104-08663}, which comprises over 17,000 unique queries in total. Each dataset reflects a different retrieval task—ranging from fact verification (e.g., FEVER), biomedical search (e.g., TREC-COVID), and scientific QA (e.g., SciFact), to community forums and entity retrieval. BEIR allows us to assess out-of-distribution generalization without additional fine-tuning, making it suitable for testing the robustness of instruction-tuned rerankers.

To further assess the reasoning capabilities of our model, we evaluate on BRIGHT~\cite{su2025brightrealisticchallengingbenchmark}, a recently introduced benchmark targeting reasoning-intensive reranking. BRIGHT contains 1,278 queries that cover factual reasoning, commonsense inference, and multi-hop question answering. Compared to BEIR, BRIGHT places greater emphasis on cognitive complexity and global consistency in ranking outputs.

For all ranking datasets, we apply a sliding window strategy~\cite{sun2023chatgpt} with window size 20 and stride 10 to address long input sequences. Candidate passages for each query are partitioned into overlapping windows, reranked independently, and recombined to produce a complete ranked list. This method balances computational tractability with the ability to rank long candidate lists under the context length constraints of transformer-based models.

Lastly, to examine whether our reranking-specific fine-tuning compromises general-purpose reasoning, we evaluate on the MMLU benchmark~\cite{hendrycks2020measuring}, which includes 57 academic and professional subject areas such as mathematics, history, law, and medicine. MMLU consists of 5,000 multiple-choice questions and is designed to measure zero-shot generalization and broad language understanding. We compare performance before and after RaCT training to assess whether the model maintains its foundational language modeling capabilities while learning specialized ranking behavior.

\subsection{Sliding Window Strategy}
\label{sliding window}
To handle long input lists during reranking, we adopt a sliding window strategy, commonly used in recent reranking models such as RankZephyr, RankGPT, and our own method. In this approach, the full list of candidate passages for each query is divided into overlapping windows—each containing a fixed number of passages (e.g., 20)—with a predefined stride (e.g., 10). Each window is treated as an independent input to the reranker, allowing the model to score passages within a manageable context size.

Due to the overlapping nature of the windows, some passages may appear in multiple segments and thus receive multiple relevance scores. However, as standard evaluation metrics like nDCG@10 only consider the top 10 ranked results, we adopt the common practice of using the final ranking produced by concatenating and scoring all passages across windows. When a passage appears more than once, its latest assigned score—i.e., the score from the last window in which it appears—is retained. Earlier scores are not explicitly reconciled. This strategy offers a simple and efficient way to apply reranking models to long passage lists while remaining consistent with prior work.

\subsection{Results}

Table~\ref{tab:mainresults} reports nDCG@10 scores for RaCT and several strong reranking baselines across four TREC Deep Learning benchmarks: DL19, DL20, DL21, and DL22, as well as the MMLU benchmark. We compare RaCT with closed-source models (RankGPT$_{3.5}$ and RankGPT$_{4}$) and open-source rerankers (RankZephyr and RankVicuna). The results demonstrate that RaCT consistently outperforms both proprietary and public systems across all ranking tasks, while maintaining competitive general-purpose reasoning capabilities.

On DL19, which contains 43 queries with high-quality human relevance annotations, RaCT achieves 75.8 in nDCG@10, outperforming RankGPT$_{4}$ (75.0), RankZephyr (74.2), and RankVicuna (66.8). The 1.6-point margin over RankZephyr highlights the effectiveness of RaCT’s stepwise chain-of-thought (CoT) mechanism in constructing globally coherent rankings. Importantly, RaCT surpasses even RankGPT$_{4}$, despite relying solely on public models and training data.

DL20 includes 54 queries and is often considered less challenging due to overlap with training sets. Nevertheless, RaCT maintains its lead with an nDCG@10 of 72.0, exceeding RankGPT$_{4}$ (70.4) and RankZephyr (70.9). These gains confirm that RaCT does not sacrifice performance in shallower reasoning scenarios, preserving accuracy on queries with lower complexity.

For DL21 and DL22—constructed from MS MARCO V2 and designed to introduce greater retrieval noise and semantic diversity—RaCT continues to perform competitively. On DL21 (66 queries), it achieves 70.6, outpeform RankZephyr (60.0) RankGPT$_{4}$ (68.4). On DL22 (75 queries), RaCT attains 53.2, leading both RankZephyr (40.8) and RankGPT$_{4}$ (50.9). These results suggest that RaCT generalizes well under distribution shift, likely due to the RPO stage encouraging preference alignment under more varied passage-query conditions.

On MMLU, a 57-domain academic QA benchmark for general reasoning ability, RaCT matches LLaMA3.1's performance, while \emph{LLaMA3.1 + SFT} shows a slight drop and RankVicuna performs poorly. This decline is expected because fine-tuning on reranking data shifts the model toward ranking-specific behaviors, which may slightly reduce performance on unrelated reasoning tasks. RankZephyr has completely lost its ability to generate meaningful outputs, receiving a score of 0 on MMLU. We also note that RankGPT-4 outperforms RaCT on this benchmark, which can be attributed to the much stronger general reasoning capability of GPT-4 as a proprietary large model, whereas RaCT is designed to enhance open-source reranking without additional inference cost.

Table~\ref{tab:beir-rerankers} presents nDCG@10 results for RaCT and baseline rerankers on nine datasets from the BEIR benchmark~\cite{DBLP:journals/corr/abs-2104-08663}, covering a range of domains including fact verification (FEVER, SciFact), open-domain QA (HotpotQA, NQ), entity-centric and financial search (DBPedia, FiQA), and scientific/biomedical corpora (Climate-FEVER, TREC-COVID). All models rerank the top 100 passages retrieved by BM25 without task-specific fine-tuning.

RaCT achieves the highest overall average (54.9), substantially outperforming RankZephyr (41.8), RankVicuna (43.3), and BM25 (40.6). This demonstrates strong zero-shot generalization, despite RaCT being trained solely on synthetic GPT-labeled data. It establishes new best results on all nine BEIR tasks.

Notably, RaCT achieves exceptional scores on reasoning-intensive datasets: 82.3 on FEVER, 79.0 on SciFact, and 76.2 on HotpotQA—   surpassing RankZephyr and RankVicuna by over 10 points in some cases. These datasets require multi-hop reasoning and alignment across evidence chains, highlighting RaCT’s advantage in tasks that go beyond surface-level lexical overlap.

On entity-heavy and domain-specific datasets such as DBPedia (45.9), NQ (60.0), and FiQA (43.5), RaCT again leads, indicating robustness across both general and technical domains. Even in low-performing and noisy retrieval settings like Climate-FEVER (23.7), SciDocs (20.7), and NFCorpus (39.2), RaCT maintains its superiority, suggesting strong relevance modeling under ambiguity and vocabulary mismatch.

\begin{figure}[ht]
\centering
\includegraphics[width=1.0\linewidth]{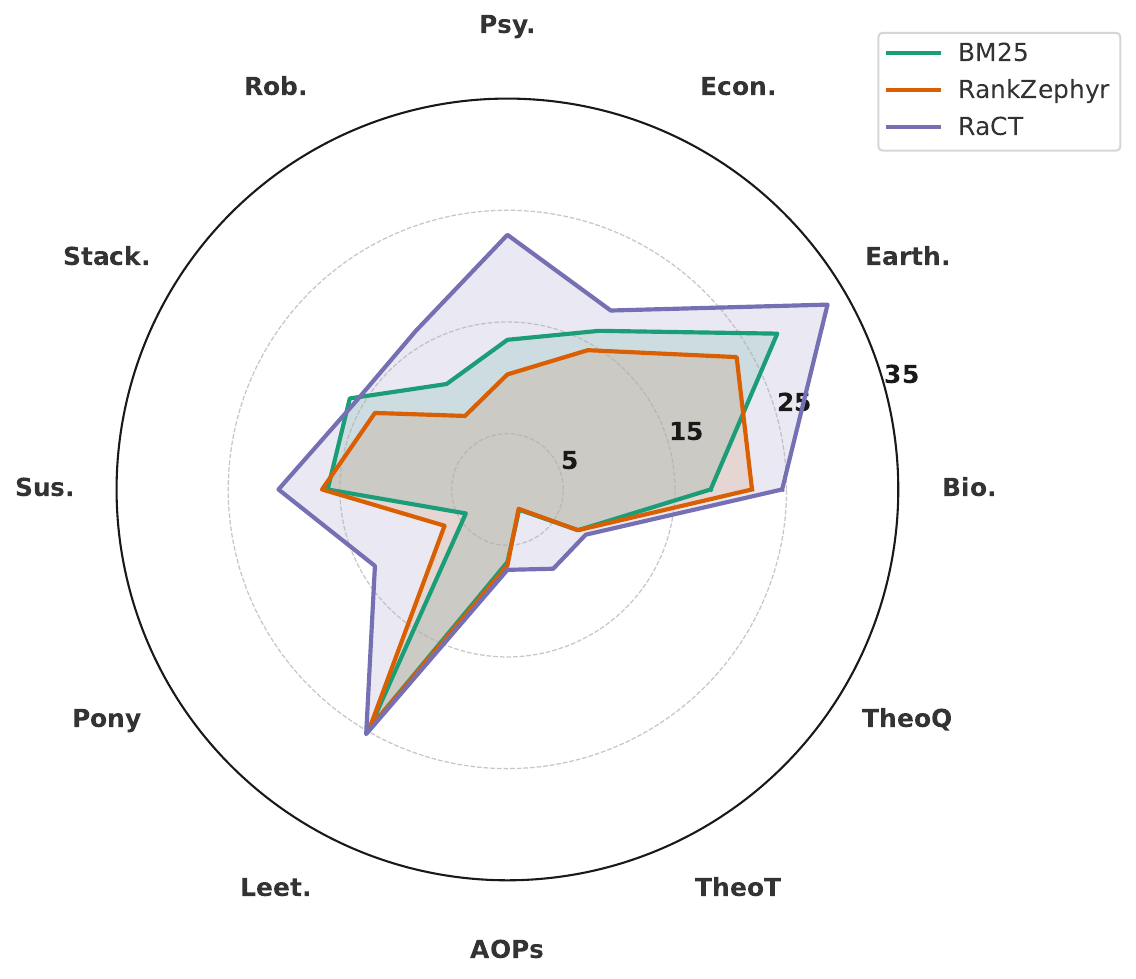}
\caption{Comparison of BM25, RankZephyr, and RaCT on the BRIGHT benchmark (nDCG@10 in \%). The radar chart highlights RaCT's superior performance across most of the 12 evaluated domains.}
\label{fig:bright_radar}
\end{figure}

These BEIR results validate RaCT’s two-stage training design. The combination of CoT supervision and preference optimization enables the model to generalize across unseen tasks, surpassing both traditional rankers and prior instruction-tuned baselines.

Figure~\ref{fig:bright_radar} offers a fine-grained view of model performance across twelve BRIGHT subsets, visualized as a radar chart. Each axis corresponds to a reasoning category such as Earth science, Psychology, or Economics. RaCT consistently dominates both BM25 and RankZephyr across nearly all categories, with the largest gains on Earth, Economy, and Psychology, which demand multi-step inference and abstract comprehension.

While BM25 remains competitive in a few domains like Leetcode, TheoT, and Stack Overflow. RaCT demonstrates significant improvements on challenging subsets such as Pony, Robotics, and Sustainble living, suggesting enhanced contextualization and decision quality. These results affirm that RaCT’s stepwise reasoning process enables generalization across diverse and cognitively demanding query types.

Taken together, the results across DL, BEIR, and reasoning subsets show that RaCT provides a scalable, robust, and accurate solution for listwise reranking—achieving state-of-the-art performance while preserving reasoning abilities across both seen and unseen evaluation settings.

\subsection{Ablation Study}
\begin{table}[h]
\centering
\begin{tabular}{l|c}
\toprule
\textbf{Setting} & \textbf{nDCG@10 (DL22)} \\
\midrule
RaCT$_\text{BM25}$ (baseline) & 0.532 \\
– CoT & 0.407 \\
– RPO & 0.504 \\
– CoT\&RPO & 0.429 \\
RaCT$_\text{SPLADE++ED}$ & 0.683 \\
\bottomrule
\end{tabular}
\caption{Ablation study with LLaMA3.1-8B as the base model. Each row removes or modifies components from the full RaCT setup and reports performance on TREC DL22 (nDCG@10).}
\label{tab:ablation}
\end{table}

Table~\ref{tab:ablation} presents a fine-grained ablation study analyzing the impact of different components in RaCT’s architecture and training pipeline. All models are based on LLaMA3.1-8B-Instruct unless otherwise noted. We examine the effect of three training stages—SFT (supervised fine-tuning), CoT (chain-of-thought supervision), and RPO (ranking preference optimization)—as well as retriever choice and model variants.

The results confirm that CoT supervision is a key contributor to performance. Enabling CoT while disabling SFT and RPO yields 75.8 on DL19 and 53.2 on DL22, substantially outperforming the SFT-only configuration (70.4/40.7). This highlights that stepwise supervision alone provides a strong inductive bias for listwise ranking. When RPO is added on top of CoT, performance further improves to 77.8 on DL19 and 68.3 on DL22 with a SPLADE++ED retriever—indicating that preference-based fine-tuning enhances ranking consistency, particularly in semantically diverse settings such as DL22.

The effect of model size is also evident. The smaller LLaMA3.2-3B-Instruct model performs worse (72.6/50.2) than its 8B counterpart under the same CoT+RPO setup, underscoring the benefits of parameter scaling. Additionally, using the base LLaMA3-8B-Instruct model results in a slight performance drop (75.5/52.2) compared to LLaMA3.1-8B-Instruct (75.8/53.2), suggesting that instruction tuning in LLaMA3.1 provides better alignment with chain-of-thought prompting, even before task-specific fine-tuning.

Overall, the ablation confirms that RaCT’s performance gains arise from the complementary contributions of instruction-tuned initialization, reasoning-aware supervision, preference-based optimization, and appropriate retriever selection.

\subsection{Inference Efficiency Analysis}

Although RaCT introduces chain-of-thought supervision during training, the inference process remains identical to that of conventional LLM-based rerankers. Specifically, RaCT outputs the final ranked list directly without generating additional intermediate steps. As a result, there is no increase in input or output token length compared to standard rerankers. We empirically measured inference efficiency using LLaMA-3.1-8B-Instruct as the base model, and observed that the average time cost per inference remains unchanged at \textbf{5.72 seconds}, which is identical to the baseline. This confirms that RaCT achieves improved ranking effectiveness without incurring additional inference overhead.

\subsection{Analysis Across Window and Stride Sizes}

Table~\ref{table: window size} examines the robustness of RaCT under varying window and stride configurations on DL19 using nDCG@10. We evaluate all models using $(2/1)$, $(10/5)$, and $(20/10)$ settings, representing increasing amounts of passage context per input.

RaCT consistently outperforms all baselines across configurations, indicating strong resilience to input segmentation. At the smallest window $(2/1)$, RaCT achieves 70.2, significantly higher than RankZephyr (56.2) and LLaMA3.1+SFT (56.7), both of which degrade due to limited context. As the window expands to $(10/5)$ and $(20/10)$, RaCT’s scores rise to 75.4 and 75.8, showing steady gains without saturation. This consistent trend reflects the benefit of RaCT’s two-stage training: CoT reasoning captures stepwise logic, while RPO enhances robustness to variations in input structure by reinforcing preference alignment across different chunk boundaries.

\begin{table}[h!]
\centering
\begin{adjustbox}{max width=\textwidth}
\begin{tabular}{l|c c c}
    \toprule
    \multirow{2}{*}{\textbf{Models}} & \multicolumn{3}{c}{\textbf{Window/Stride Size}} \\
    \cmidrule(lr){2-4}
    & \textbf{2/1} & \textbf{10/5} & \textbf{20/10} \\
    \midrule
    Gemma-7B  & 54.1 & 59.3 & 53.3\\
    LLaMA3.1  & 54.9 & 66.0 & 64.1 \\
    RankZephyr  & 56.2 & 61.2 & 74.2\\
    RankVicuna  & 55.1 & 65.5 & 66.8  \\
    LLaMA3.1 + SFT & 56.7 & 66.3 & 66.6  \\
    RaCT w/o RPO & 69.7 & 73.7 & 75.3 \\
    RaCT & 70.2 & 75.4 & 75.8 \\
    \bottomrule
\end{tabular}
\end{adjustbox}
\caption{Performance of different models with different window size and step size, BM25 performs all of the retrieval stage, and all of the evaluations are performed on DL19}
\label{table: window size}
\end{table}

In contrast, other models show more volatility. RankZephyr, for example, improves by over 18 points from $(2/1)$ to $(20/10)$, revealing greater dependence on broader input windows to offset reasoning limitations. RaCT’s stable and high performance across window sizes shows that it generalizes better under shifting segmentation constraints, making it more adaptable for applications with strict token budgets or latency requirements.

\subsection{Qualitative Comparison on General Prompting}

To assess how well reranking models generalize beyond their primary task, we conduct a qualitative comparison using a general-purpose prompt: \emph{"Compose an engaging travel blog post about a recent trip to Hawaii, highlighting cultural experiences and must-see attractions."}. Figure \ref{fig:hawaii-case-study} presents the responses from RankVicuna,
RankZephyr, and RaCT.

While RankZephyr performs well in reranking by producing a valid ranked list, it lacks flexibility in open-ended generation. Faced with a creative prompt, it rigidly outputs ranked passage IDs, indicating limited adaptability beyond its reranking training. Despite instruction tuning, it does not generalize well when the task format diverges from its training distribution.

In contrast, RaCT generates a coherent travel blog post with clear structure, meaningful content sections, and actionable tips. This response adheres closely to the prompt and showcases RaCT’s capacity for zero-shot, instruction-following generation. RankVicuna provides a brief summary and plausible title but lacks detail and usefulness.

These results highlight the limitations of reranking-only tuning and underscore RaCT’s broader generalization ability, enabled by its CoT reasoning and two-stage training approach.

\section{Conclusion and Future Work}

We presented RaCT, a zero-shot listwise reranker built on LLaMA3.1 that frames reranking as a chain-of-thought reasoning task. Through two-stage training—supervised CoT fine-tuning followed by ranking preference optimization—RaCT achieves strong performance across standard and out-of-distribution benchmarks. It outperforms both open-source baselines such as RankZephyr and closed-source models like RankGPT$_{4}$, while retaining general reasoning ability.

In future work, we plan to extend RaCT to additional architectures, including Mistral, Zephyr, and QWEN, and incorporate more diverse and high-quality training data to improve robustness. We also aim to explore alignment strategies that better support multi-objective reasoning tasks beyond document ranking.


\begin{figure}[t]
\centering

\begin{tcolorbox}[colframe=blue!50!black, colback=blue!10!white, title=Prompt, width=\columnwidth, boxrule=0.5mm, sharp corners]
    Compose an engaging travel blog post about a recent trip to Hawaii, highlighting cultural experiences and must-see attractions.
\end{tcolorbox}
\begin{tcolorbox}[colframe=red!50!black, colback=red!5!white, title=RankVicuna , width=\columnwidth, boxrule=0.5mm, sharp corners]
    \textbf{Ultimately, showcasing Hawaii's rich history and cultural significance.} \\
    \textbf{Title:} Embracing Hawaiian Culture: Discovering Historical Landmarks and Rich Traditions on the Big Island
\end{tcolorbox}

\begin{tcolorbox}[colframe=green!50!black, colback=green!5!white, title=RankZephyr , width=\columnwidth, boxrule=0.5mm, sharp corners]
    [1] > [2] > [3] > [4] > [5] > [6] > [7] > [8] > [9] > [10] > [2] > [11] > [12] > [13]
\end{tcolorbox}

\begin{tcolorbox}[colframe=orange!50!black, colback=orange!5!white, title=RaCT , width=\columnwidth, boxrule=0.5mm, sharp corners]
    \textbf{Aloha and Mahalo: Discovering the Aloha Spirit in Hawaii} \\
    As I stepped off the plane in Honolulu, I couldn't help but feel a sense of excitement and anticipation wash over me. I had dreamed of visiting Hawaii for years, and finally, I was here. The warm tropical air enveloped me, and the sweet scent of plumeria filled my senses as I made my way to my hotel. Little did I know, this trip would be a journey not only to breathtaking natural wonders but also to a rich cultural heritage that would leave me feeling grateful and inspired.

    \textbf{Immersing in the Local Culture:} \\
    \emph{(Content omitted for brevity.)}

    \textbf{Must-See Attractions:} \\
    \emph{(Content omitted for brevity.)}

    \textbf{A Taste of Aloha:} \\
    \emph{(Content omitted for brevity.)}

    \textbf{Conclusion:} \\
    \emph{(Content omitted for brevity.)}

    \textbf{Practical Tips and Recommendations:}\\
    \emph{(Content omitted for brevity.)}\\
    Mahalo (thank you) for joining me on this journey to the Aloha State. I hope my story has inspired you to plan your own trip to Hawaii.
\end{tcolorbox}

\caption{Qualitative comparison on a travel blog prompt. RankZephyr outputs a correct ranked list but struggles with open-ended generation. RaCT produces a coherent, structured response, showing better generalization and instruction-following.}
\label{fig:hawaii-case-study}
\end{figure}

\clearpage

\bibliographystyle{ACM-Reference-Format}
\balance
\bibliography{main}

\end{document}